\PassOptionsToPackage{prologue,table}{xcolor}
\documentclass[sigconf]{acmart}

\copyrightyear{2024}
\acmYear{2024}
\setcopyright{acmlicensed}\acmConference[ICSE-Companion '24]{2024 IEEE/ACM 46th International Conference on Software Engineering: Companion Proceedings}{April 14--20, 2024}{Lisbon, Portugal}
\acmBooktitle{2024 IEEE/ACM 46th International Conference on Software Engineering: Companion Proceedings (ICSE-Companion '24), April 14--20, 2024, Lisbon, Portugal}
\acmDOI{10.1145/3639478.3643086}
\acmISBN{979-8-4007-0502-1/24/04}

\AtBeginDocument{%
  }

\usepackage[table]{xcolor}
\usepackage{graphicx, subfigure}
\usepackage{caption}
\usepackage{makecell}
\usepackage{soul}
\usepackage{listings}
\usepackage{multirow}
\usepackage{pifont}
\usepackage{xspace}
\usepackage{epigraph} 
\usepackage[flushleft]{threeparttable}
\usepackage{url}
\definecolor{backcolour}{rgb}{0.95,0.95,0.92}
\usepackage{algorithm2e}
\UseRawInputEncoding
\usepackage[colorinlistoftodos]{todonotes}
\usepackage{listings}
\usepackage{comment}
\usepackage{fontawesome}

\definecolor{celestialblue}{rgb}{0.29, 0.59, 0.82}
\definecolor{awesome}{rgb}{0.0, 0.2, 0.6}
\definecolor{coolblack}{rgb}{0.0, 0.18, 0.39}
\definecolor{maroon}{cmyk}{0, 0.87, 0.68, 0.32}
\definecolor{halfgray}{gray}{0.55}
\definecolor{ipython_frame}{RGB}{207, 207, 207}
\definecolor{ipython_bg}{RGB}{247, 247, 247}
\definecolor{ipython_red}{RGB}{186, 33, 33}
\definecolor{ipython_green}{RGB}{0, 128, 0}
\definecolor{ipython_cyan}{RGB}{64, 128, 128}
\definecolor{ipython_purple}{RGB}{170, 34, 255}
\usepackage[strict]{changepage}
  \definecolor{ABlue}{HTML}{127bca}
 \definecolor{LHScolor}{HTML}{555555}
\usepackage{framed}

\definecolor{formalshade}{rgb}{1.0,1.0,1.0}
\definecolor{side}{rgb}{0.0,0.2,0.6}

\definecolor{gray(x11gray)}{rgb}{0.75, 0.75, 0.75}

\lstdefinelanguage{python}{
    morekeywords={access,and,break,class,continue,def,del,elif,else,except,exec,finally,for,from,global,if,import,in,is,lambda,not,or,pass,print,raise,return,try,while},
    morekeywords=[2]{abs,all,any,basestring,bin,bool,bytearray,callable,chr,classmethod,cmp,compile,complex,delattr,dict,dir,divmod,enumerate,eval,execfile,file,filter,float,format,frozenset,getattr,globals,hasattr,hash,help,hex,id,input,int,isinstance,issubclass,iter,len,list,locals,long,map,max,memoryview,min,next,object,oct,open,ord,pow,property,range,raw_input,reduce,reload,repr,reversed,round,set,setattr,slice,sorted,staticmethod,str,sum,super,tuple,type,unichr,unicode,vars,xrange,zip,apply,buffer,coerce,intern},
    sensitive=true,
    morecomment=[l]\#,
    morestring=[b]',
    morestring=[b]",
    morestring=[s]{'''}{'''},
    morestring=[s]{"""}{"""},
    morestring=[s]{r'}{'},
    morestring=[s]{r"}{"},
    morestring=[s]{r'''}{'''},
    morestring=[s]{r"""}{"""},
    morestring=[s]{u'}{'},
    morestring=[s]{u"}{"},
    morestring=[s]{u'''}{'''},
    morestring=[s]{u"""}{"""},
    literate=
    {á}{{\'a}}1 {é}{{\'e}}1 {í}{{\'i}}1 {ó}{{\'o}}1 {ú}{{\'u}}1
    {Á}{{\'A}}1 {É}{{\'E}}1 {Í}{{\'I}}1 {Ó}{{\'O}}1 {Ú}{{\'U}}1
    {à}{{\`a}}1 {è}{{\`e}}1 {ì}{{\`i}}1 {ò}{{\`o}}1 {ù}{{\`u}}1
    {À}{{\`A}}1 {È}{{\'E}}1 {Ì}{{\`I}}1 {Ò}{{\`O}}1 {Ù}{{\`U}}1
    {ä}{{\"a}}1 {ë}{{\"e}}1 {ï}{{\"i}}1 {ö}{{\"o}}1 {ü}{{\"u}}1
    {Ä}{{\"A}}1 {Ë}{{\"E}}1 {Ï}{{\"I}}1 {Ö}{{\"O}}1 {Ü}{{\"U}}1
    {â}{{\^a}}1 {ê}{{\^e}}1 {î}{{\^i}}1 {ô}{{\^o}}1 {û}{{\^u}}1
    {Â}{{\^A}}1 {Ê}{{\^E}}1 {Î}{{\^I}}1 {Ô}{{\^O}}1 {Û}{{\^U}}1
    {œ}{{\oe}}1 {Œ}{{\OE}}1 {æ}{{\ae}}1 {Æ}{{\AE}}1 {ß}{{\ss}}1
    {ç}{{\c c}}1 {Ç}{{\c C}}1 {ø}{{\o}}1 {å}{{\r a}}1 {Å}{{\r A}}1
    {€}{{\EUR}}1 {£}{{\pounds}}1
    {^}{{{\color{ipython_purple}\^{}}}}1
    {=}{{{\color{ipython_purple}=}}}1
    {+}{{{\color{ipython_purple}+}}}1
    {*}{{{\color{ipython_purple}$^\ast$}}}1
    {/}{{{\color{ipython_purple}/}}}1
    {+=}{{{+=}}}1
    {-=}{{{-=}}}1
    {*=}{{{$^\ast$=}}}1
    {/=}{{{/=}}}1,
    literate=
    *{-}{{{\color{ipython_purple}-}}}1
     {?}{{{\color{ipython_purple}?}}}1,
    identifierstyle=\color{black}\ttfamily,
    commentstyle=\color{ipython_cyan}\ttfamily,
    stringstyle=\color{ipython_red}\ttfamily,
    keepspaces=true,
    showspaces=false,
    showstringspaces=false,
    rulecolor=\color{ipython_frame},
    numberstyle=\tiny\color{halfgray},
    backgroundcolor=\color{ipython_bg},
    basicstyle=\scriptsize,
    keywordstyle=\color{ipython_green}\ttfamily,
}

\definecolor{chestnut}{rgb}{0.8, 0.36, 0.36}

\definecolor{chestnut}{rgb}{0.8, 0.36, 0.36}

\usepackage[skins,breakable]{tcolorbox}

\newcommand{\caseOne}{Colors.js}
\newcommand{\caseTwo}{es5-ext}
\newcommand{\caseThree}{Ua-parser}

\definecolor{Large}{HTML}{696969}
\definecolor{Negligible}{HTML}{D3D3D3}
\definecolor{Medium}{HTML}{808080}
\definecolor{Small}{HTML}{A9A9A9}

\begin{document}

\title{Going Viral: Case Studies on the Impact of Protestware}

\author{Youmei Fan}
\affiliation{%
  \country{Nara Institute of Science and Technology, Japan}
}
\email{fan.youmei.fs2@is.naist.jp}

\author{Dong Wang}
\affiliation{%
  \institution{Tianjin University}
  \country{China}
}
\email{d.wang@ait.kyushu-u.ac.jp}

\author{Supatsara Wattanakriengkrai}
\affiliation{%
  \country{Nara Institute of Science and Technology, Japan}
}
\email{wattanakri.supatsara.ws3@is.naist.jp}

\author{Hathaichanok Damrongsiri}
\affiliation{%
  \country{Nara Institute of Science and Technology, Japan}
}
\email{damrongsiri.hathaichanok.db5@is.naist.jp}

\author{Christoph Treude}
\affiliation{%
  \institution{Singapore Management University}
  \country{Singapore}
}
\email{ctreude@smu.edu.sg}

\author{Hideaki Hata}
\affiliation{%
  \institution{Shinshu University}
  \country{Japan}
}
\email{hata@shinshu-u.ac.jp}

\author{Raula Gaikovina Kula}
\affiliation{%
  \country{Nara Institute of Science and Technology, Japan}
}
\email{raula-k@is.naist.jp}

\renewcommand{\shortauthors}{Fan et al.}

\begin{abstract}
Maintainers are now self-sabotaging their work in order to take political or economic stances, a practice referred to as ``protestware''.
In this poster, we present our approach to understand how the discourse about such an attack went viral, how it is received by the community, and whether developers respond to the attack in a timely manner. 
We study two notable protestware cases, i.e., \caseOne~and \caseTwo, comparing with discussions of a typical security vulnerability as a baseline, i.e., \caseThree, and perform a thematic analysis of more than two thousand protest-related posts to extract the different narratives when discussing protestware.


\end{abstract}



\maketitle
\epigraph{
\color{coolblack}{\textit{``In dev-land, we don't stand on the shoulders of giants. We keep our life-rafts afloat by sticky-taping together skerricks of code that hopefully has more buoyancy than ballast. And sometime it just takes one person to take the whole ship down.'' }}}{\textit{commentary} on protestware \faGithub}
\vspace{-1em}
\epigraph{
\color{coolblack}{\textit{``Anyone who experienced actual significant disruption from this (protestware) brought it on themselves with their bad dev practices. No one forced anyone to install the latest version without actually verifying it at all. Didnt corrupt the version history so nope, people just letting their entitlement and lack of understanding of licenses show.'' }}}{\textit{commentary} on protestware \faGithub}
\vspace{-2em}
\section{Extended Abstract}
Open source software development has emerged as an unexpected platform for expressing social and political protests. In recent years, we have witnessed the emergence of ``protestware'', defined as software that developers deliberately modify to express dissent and draw attention to issues they consider important~\cite{kula2022war}. Such protests can have significant social and technical impacts, especially given the interconnected nature of modern software systems.

Protestware takes on many forms. For example, the maintainer of `node-ipc' used malicious code to target host machines with IP addresses in Russia or Belarus in response to the War in Ukraine~\cite{massacci2022free}. 
GitHub declared this a critical vulnerability, known as CVE-2022-23812~\cite{CVE-2022}.
This act underscores the political motivations that can inspire the creation of protestware. Other instances of protestware stem from industry concerns, such as the perceived exploitation of open-source labor by large corporations. For instance, the developer of the `faker' library intentionally introduced an infinite loop, disrupting thousands of projects~\cite{bellovin2022open}. Today, `faker' is a community-controlled project, maintained by a team of developers from various companies~\cite{fakerjsUpdateFrom}.
Such protests can have widespread consequences, particularly since modern software systems often rely heavily on third-party libraries, making them potential points of vulnerability~\cite{zahan2022weak}. To illustrate the scale of this interconnectedness, libraries listed in the popular NPM registry, which hosts more than a million libraries, each depend on an average of five to six other libraries within the same ecosystem~\cite{chinthanet2021makes}.
Beyond technical issues, the rise of protestware threatens the trust that is foundational to modern software ecosystems~\cite{ghofrani2022trust}, creating disruptions within the software development community. Developers using a protesting library must decide whether to continue using it or take on the potentially complex task of finding alternatives~\cite{he2021multi} or reverting to older versions~\cite{kula2018developers}. Societal reactions to protestware vary, with responses ranging from strong support to harsh criticism, and everything in between~\cite{massacci2022free}.

\vspace{-1em}
\paragraph{\textbf{Motivation}}
The motivation that will be illustrated by this poster is to track and characterize disruptions caused by protestware and the immediate reactions it prompted.
Concretely, we would like to understand the extent to which the software ecosystem is affected by discussions of the protestware, and what is the nature of the discussions. 

    
\vspace{-0.5em}
\paragraph{\textbf{Approach}}

In this study, we apply a comprehensive empirical study approach by combining quantitative analysis and later a systematic manual coding of the different protestware discussions. 
We conduct a large-scale empirical study by mining both the GitHub platform and social media (i.e., Twitter, Reddit, HackerNews) to extract discussions of the two case studies of protestware. 
We then conduct two analyses.

First, a quantitative analysis is conducted, focusing on the abandonment of protestware by developers. The approach involves identifying valid JavaScript repositories using protestware as a dependency, mining historical dependency changes, and retrieving popularity metrics using the GitHub API. The results reveal a lower likelihood that developers abandon protestware, with specific insights into the characteristics of influenced repositories and the distribution of the changes in dependencies. The spread of protestware discussions throughout the ecosystem is explored through two analyses, quantifying the speed of spread, and visualizing the evolution of disruption across GitHub. 

For the second analysis, we extract characteristics of posts related to protestware using manual coding. We categorize posts into opinion-related and technical categories, classify opinion-related posts to identify the protestware stance, and examine technical posts related to mitigation. The overall methodology integrates both quantitative and qualitative approaches, providing a comprehensive understanding of the dynamics surrounding protestware and its impact on the developer community.
\vspace{-0.5em}
\paragraph{\textbf{Key results so far}}
Our results indicate that both cases of protestware (\caseOne~and \caseTwo) have comparable disruption to that of a security vulnerability (i.e., \caseThree). However, unlike a security vulnerability, developers are less likely to abandon the dependency. For example, of the 146 repositories that discussed the \caseOne~issue, only 15\% abandoned the dependency, compared to 82\% by dependents of \caseThree.
Measuring the spread, we find that the news of protestware can be broken through social media outlets (i.e., \caseOne), and may have a greater reach in terms of developers and repositories (\caseOne) compared to \caseThree, taking two days less to reach 2,000 posts, and almost 53 days less to reach 10,000 developers. Additionally, it should be noted that Twitter seemed to break the news of the protest, even before the original issue was created on GitHub. For example, the first tweet,\footnote{\url{https://twitter.com/andrewmcodes/status/1479876979151106048}} directly asked the protester about the injected protest code.

By categorizing posts into opinions and technical discussions, we show that \caseOne~ demonstrated a mixture of opinions (50.3\%) and technical discussions (44.6\%). Developers impacted by \caseTwo, on the other hand, predominantly engaged in expressing their opinions (77.3\%) rather than focusing on technical aspects (22.6\%). 
Furthermore,  we identified four main themes in the opinions expressed during the discussions: \textit{1. OSS Philosophy}, \textit{2. Legal Issues and Rights}, \textit{3. Trust and Reputation}, and \textit{4. Supportive or Dogmatic responses}.

\paragraph{\textbf{Looking Ahead}}
At this early stage, it is unsure whether or not protestware is here to stay in the open source software ecosystem. 
With the heavy reliance on the ecosystem and its supply chain, the mere presence of protestware is a reminder that software development is becoming increasingly intertwined with society and its issues, ranging from economic to political. 
In this first large-scale exploration into protestware, we present the different narratives that drive this emerging phenomenon. 
These future research opportunities not only include providing developers with insight into varying perspectives but also offer actionable guidance to foster constructive dialogue and reduce toxicity in these sensitive discussions. 
Such frameworks could transform protestware from a potential point of contention into an instrument for positive social change, aligning the open-source community with broader societal values and goals.
\vspace{-1em}

\section*{ACKNOWLEDGEMENTS}
This work was supported by JST SICORP Grant Number JPMJSC2206.

\bibliographystyle{ACM-Reference-Format}
\bibliography{acmart}


\begin{thebibliography}{10}


\ifx \showCODEN    \undefined \def \showCODEN     #1{\unskip}     \fi
\ifx \showDOI      \undefined \def \showDOI       #1{#1}\fi
\ifx \showISBNx    \undefined \def \showISBNx     #1{\unskip}     \fi
\ifx \showISBNxiii \undefined \def \showISBNxiii  #1{\unskip}     \fi
\ifx \showISSN     \undefined \def \showISSN      #1{\unskip}     \fi
\ifx \showLCCN     \undefined \def \showLCCN      #1{\unskip}     \fi
\ifx \shownote     \undefined \def \shownote      #1{#1}          \fi
\ifx \showarticletitle \undefined \def \showarticletitle #1{#1}   \fi
\ifx \showURL      \undefined \def \showURL       {\relax}        \fi
\providecommand\bibfield[2]{#2}
\providecommand\bibinfo[2]{#2}
\providecommand\natexlab[1]{#1}
\providecommand\showeprint[2][]{arXiv:#2}

\bibitem[fak({[n.\,d.]})]%
        {fakerjsUpdateFrom}
 \bibinfo{year}{[n.\,d.]}\natexlab{}.
\newblock \bibinfo{title}{{A}n update from the {F}aker team | {F}aker --- fakerjs.dev}.
\newblock \bibinfo{howpublished}{\url{https://fakerjs.dev/about/announcements/2022-01-14.html}}.
\newblock
\newblock
\shownote{[Accessed 20-Jul-2023]}.


\bibitem[CVE(2022)]%
        {CVE-2022}
 \bibinfo{year}{2022}\natexlab{}.
\newblock \bibinfo{booktitle}{\emph{CVE-2022-23812}}.
\newblock
\urldef\tempurl%
\url{https://nvd.nist.gov/vuln/detail/cve-2022-23812}
\showURL{%
\tempurl}


\bibitem[Bellovin(2022)]%
        {bellovin2022open}
\bibfield{author}{\bibinfo{person}{Steven~M Bellovin}.} \bibinfo{year}{2022}\natexlab{}.
\newblock \showarticletitle{Open Source and Trust}.
\newblock \bibinfo{journal}{\emph{IEEE Security \& Privacy}} \bibinfo{volume}{20}, \bibinfo{number}{02} (\bibinfo{year}{2022}), \bibinfo{pages}{107--108}.
\newblock


\bibitem[Chinthanet et~al\mbox{.}(2021)]%
        {chinthanet2021makes}
\bibfield{author}{\bibinfo{person}{Bodin Chinthanet}, \bibinfo{person}{Brittany Reid}, \bibinfo{person}{Christoph Treude}, \bibinfo{person}{Markus Wagner}, \bibinfo{person}{Raula~Gaikovina Kula}, \bibinfo{person}{Takashi Ishio}, {and} \bibinfo{person}{Kenichi Matsumoto}.} \bibinfo{year}{2021}\natexlab{}.
\newblock \showarticletitle{What makes a good Node. js package? Investigating Users, Contributors, and Runnability}.
\newblock \bibinfo{journal}{\emph{arXiv preprint arXiv:2106.12239}} (\bibinfo{year}{2021}).
\newblock


\bibitem[Ghofrani et~al\mbox{.}(2022)]%
        {ghofrani2022trust}
\bibfield{author}{\bibinfo{person}{Javad Ghofrani}, \bibinfo{person}{Paria Heravi}, \bibinfo{person}{Kambiz~A Babaei}, {and} \bibinfo{person}{Mohammad~D Soorati}.} \bibinfo{year}{2022}\natexlab{}.
\newblock \showarticletitle{Trust challenges in reusing open source software: An interview-based initial study}. In \bibinfo{booktitle}{\emph{Proceedings of the 26th ACM International Systems and Software Product Line Conference-Volume B}}. \bibinfo{pages}{110--116}.
\newblock


\bibitem[He et~al\mbox{.}(2021)]%
        {he2021multi}
\bibfield{author}{\bibinfo{person}{Hao He}, \bibinfo{person}{Yulin Xu}, \bibinfo{person}{Yixiao Ma}, \bibinfo{person}{Yifei Xu}, \bibinfo{person}{Guangtai Liang}, {and} \bibinfo{person}{Minghui Zhou}.} \bibinfo{year}{2021}\natexlab{}.
\newblock \showarticletitle{A multi-metric ranking approach for library migration recommendations}. In \bibinfo{booktitle}{\emph{2021 IEEE International Conference on Software Analysis, Evolution and Reengineering (SANER)}}. IEEE, \bibinfo{pages}{72--83}.
\newblock


\bibitem[Kula et~al\mbox{.}(2018)]%
        {kula2018developers}
\bibfield{author}{\bibinfo{person}{Raula~Gaikovina Kula}, \bibinfo{person}{Daniel~M German}, \bibinfo{person}{Ali Ouni}, \bibinfo{person}{Takashi Ishio}, {and} \bibinfo{person}{Katsuro Inoue}.} \bibinfo{year}{2018}\natexlab{}.
\newblock \showarticletitle{Do developers update their library dependencies? An empirical study on the impact of security advisories on library migration}.
\newblock \bibinfo{journal}{\emph{Empirical Software Engineering}}  \bibinfo{volume}{23} (\bibinfo{year}{2018}), \bibinfo{pages}{384--417}.
\newblock


\bibitem[Kula and Treude(2022)]%
        {kula2022war}
\bibfield{author}{\bibinfo{person}{Raula~Gaikovina Kula} {and} \bibinfo{person}{Christoph Treude}.} \bibinfo{year}{2022}\natexlab{}.
\newblock \showarticletitle{In war and peace: the impact of world politics on software ecosystems}. In \bibinfo{booktitle}{\emph{Proceedings of the 30th ACM Joint European Software Engineering Conference and Symposium on the Foundations of Software Engineering}}. \bibinfo{pages}{1600--1604}.
\newblock


\bibitem[Massacci et~al\mbox{.}(2022)]%
        {massacci2022free}
\bibfield{author}{\bibinfo{person}{Fabio Massacci}, \bibinfo{person}{Antonino Sabetta}, \bibinfo{person}{Jelena Mirkovic}, \bibinfo{person}{Toby Murray}, \bibinfo{person}{Hamed Okhravi}, \bibinfo{person}{Mohammad Mannan}, \bibinfo{person}{Anderson Rocha}, \bibinfo{person}{Eric Bodden}, {and} \bibinfo{person}{Daniel~E Geer}.} \bibinfo{year}{2022}\natexlab{}.
\newblock \showarticletitle{“Free” as in Freedom to Protest?}
\newblock \bibinfo{journal}{\emph{IEEE Security \& Privacy}} \bibinfo{volume}{20}, \bibinfo{number}{5} (\bibinfo{year}{2022}), \bibinfo{pages}{16--21}.
\newblock


\bibitem[Zahan et~al\mbox{.}(2022)]%
        {zahan2022weak}
\bibfield{author}{\bibinfo{person}{Nusrat Zahan}, \bibinfo{person}{Thomas Zimmermann}, \bibinfo{person}{Patrice Godefroid}, \bibinfo{person}{Brendan Murphy}, \bibinfo{person}{Chandra Maddila}, {and} \bibinfo{person}{Laurie Williams}.} \bibinfo{year}{2022}\natexlab{}.
\newblock \showarticletitle{What are weak links in the npm supply chain?}. In \bibinfo{booktitle}{\emph{Proceedings of the 44th International Conference on Software Engineering: Software Engineering in Practice}}. \bibinfo{pages}{331--340}.
\newblock


\end{thebibliography}

\end{document}